\documentclass[aps,prl,floatfix,12pt,superscriptaddress,reprint]{revtex4-1}

\usepackage{verbatim}
\usepackage{graphicx}
\usepackage{color}
\usepackage{amsmath}
\usepackage{amssymb}
\usepackage{braket}
\usepackage{subfig}
\usepackage[hidelinks]{hyperref}

\usepackage{float}
\usepackage{siunitx}
\DeclareSIUnit{\sqrthz}{\ensuremath{\text{\hertz}^{-1/2}}}
\DeclareSIUnit{\sqrtf}{\ensuremath{\text{f}^{-1/2}}}

\begin{document}

\title[]{Digital laser frequency and intensity stabilization  based on the STEMlab platform (originally Red Pitaya)}
% Force line breaks with \\

\author{T. Preuschoff}
 \email{apqpub@physik.tu-darmstadt.de}
\author{M. Schlosser}
\author{G. Birkl}
\homepage{http://www.iap.tu-darmstadt.de/apq}
\affiliation{Institut f\"ur Angewandte Physik, Technische Universit\"at Darmstadt, Schlossgartenstr. 7, 64289 Darmstadt, Germany
}

\date{\today}

\begin{abstract}
We report on the development, implementation, and characterization of digital controllers for laser frequency stabilization as well as intensity stabilization and control. Our design is based on the STEMlab (originally Red Pitaya) platform. The presented analog hardware interfaces  provide all necessary functionalities for the designated applications and can be integrated in standard 19-inch rack mount units. Printed circuit board layouts are made available as an open-source project.\cite{APQGit_Lockbox,APQGit_IntStab} A detailed characterization shows that the bandwidth (\SI{1.25}{\mega\hertz}) and the noise performance of the controllers are limited by the STEMlab system and not affected by the supplementary hardware. Frequency stabilization of a diode laser system resulting in a linewidth of \SI{52(1)}{\kilo\hertz} (FWHM) is demonstrated. Intensity control to the \SI{1e-3}{} level with sub-microsecond rise and fall times based on an acousto-optic modulator as actuator is achieved.\\

\bfseries{This article appeared in Rev. Sci. Instrum. 91, 083001 (2020) and may be found at \href{https://doi.org/10.1063/5.0009524}{https://doi.org/10.1063/5.0009524}. It may be downloaded for personal use only. Any other use requires prior permission of the author and AIP Publishing. }
\end{abstract}

\maketitle

\section{Introduction}
Complex experiments in quantum optics and photonics often involve a multitude of laser systems that are actively stabilized in frequency and intensity. For such applications, digital controllers are favorable since they offer a wide range of control parameters without hardware modifications. The quality of the engaged control electronics is crucial for the performance of the entire laser system.

We present versatile and cost efficient digital controllers for laser frequency and intensity stabilization.\cite{APQGit_Lockbox,APQGit_IntStab} Both architectures are based on the STEMlab 125-14 board.\cite{RedPitaya} The STEMlab platform has already been successfully applied to control tasks in optical experiments, such as optical phase locking, \cite{Olaya2016,Tourigny2018} laser frequency comb stabilization,\cite{Shaw2019} second harmonic generation,\cite{Hannig2018} or as a lock-in amplifier.\cite{Stimpson2018} In our work, the STEMlab hardware is embedded in two different analog interfaces that facilitate the application to a considerable variety of laser systems. We provide a characterization of both systems. As performance benchmarks, frequency stabilization of an external cavity diode laser system (ECDL) and intensity control based on an acousto-optic modulator (AOM) as actuator are presented.

\section{Hardware design}
The design comprises ready-to-use open-source printed circuit board (PCB) layouts\cite{APQGit_Lockbox,APQGit_IntStab} that can be integrated in standard 19-inch rack mount units. The STEMlab board is directly mounted on the PCB which contains a step-down regulator (Linear Technology LT8610) for its power supply. The STEMlab system includes two fast 14-bit analog-to-digital converters (ADC) and two fast 14-bit digital-to-analog converters (DAC) controlled by a field programmable gate array (FPGA). In order to obtain optimal performance of the DAC outputs, it is essential to deactivate the noisy internal DC offset circuits.\cite{Hannig2018,Offset} This shifts the output voltage range from $\pm$\SI{1}{\volt} to 0\,--\,\SI{2}{\volt}. The STEMlab hardware is programmed using the open-source software package PyRPL\cite{PyRPL} that offers a Python application programming interface (API) with a configurable graphical user interface (GUI) and customized FPGA code. The hardware presented in this work is compatible with other software projects for programming the STEMlab platform, such as the original software \cite{RedPitaya} or a toolkit\cite{Luda2019} which offers similar functionalities as PyRPL using a web interface. Via PyRPL, up to three digital proportional-integral controllers (PI) can be implemented simultaneously on the FPGA. We observe a \SI{200(5)}{\nano\second} delay for an open-loop step response with unity proportional voltage gain (P-gain). This delay limits the control bandwidth to \SI{1.25}{\mega\hertz} (\num{\pi/2} phase margin).

\begin{figure}[htb]
	\includegraphics[width = \columnwidth]{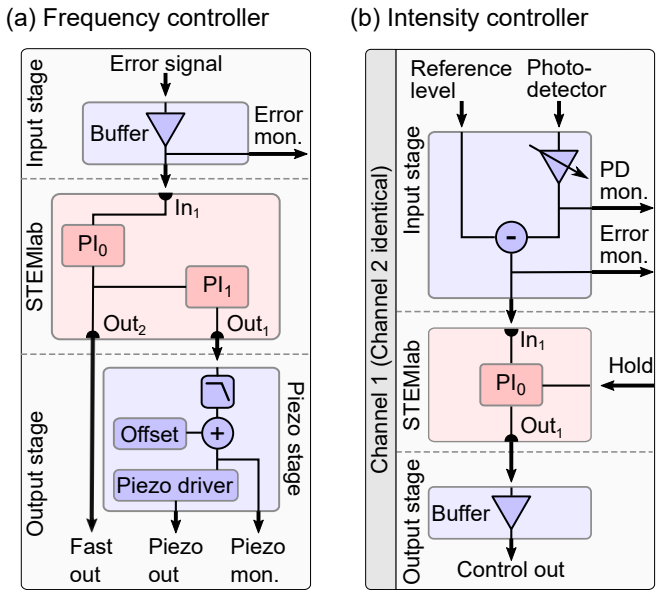}
	\caption{\label{fig:block}Block diagrams of controllers for (a) laser frequency stabilization and (b) laser intensity stabilization and control.}
\end{figure}

Figure~\ref{fig:block}(a) shows the block diagram of the controller for laser frequency stabilization. The control-loop error signal is fed into an input stage that buffers the signal and generates an additional monitoring output. The input buffer can be used as a variable amplifier or attenuator in order to match the error-signal amplitude to the ADC voltage-range (\SI{\pm 1}{\volt}) if necessary. The input stage is connected to the STEMlab board (In${}_1$). Via PyRPL, the following locking scheme is implemented: A first PI (PI${}_0$)   directly drives the fast output of the controller via one of the DACs (Out$_{2}$). The signal acts on a fast actuator, e.g. the laser current. This PI is used to stabilize the laser frequency to the error-signal zero crossing. For high-accuracy frequency stabilization it is necessary to reduce the effect of the finite DAC resolution by sufficient attenuation of the output signal at the actuator input. This restricts the control range of the fast actuator. In order to compensate for slow drifts with large amplitude, otherwise leading to saturation of the fast output, a second control loop is implemented: The output of the first PI (PI${}_0$) also serves as input for a second PI (PI${}_1$) that retains the mean output of PI${}_0$ at zero by acting on a slow actuator with large control range. For this purpose, its output  (Out${}_1$) is connected to a low-noise piezo stage driving a slow piezo actuator. In the piezo stage, the low-pass filtered signal (cut-off frequency \SI{10}{\kilo\hertz}) is amplified and offset in order to operate over the full voltage range of the analog interface ($\pm$\SI{10}{\volt}). A \SI{250}{\milli\ampere} output-buffer (Texas Instruments BUF634) allows for driving large-capacitance piezo actuators. The piezo output series resistance of \SI{4.7}{\ohm} forms an additional low-pass filter with the capacitance of the piezo.
\footnote{Piezo capacitance for the laser systems used in this work: $\sim$\SI{2.6}{\micro\farad}}

Two independent laser intensity controllers are implemented in one device using one STEMlab 125-14 board for both channels. The block diagram of Fig.~\ref{fig:block}(b) shows one channel of this device. The input stage is designed to compare a photodetector voltage with an external reference level. A stepwise-variable amplification (1,2,4,8,16) can be selected for the photodetector input. The reference signal and the amplified photodetector signal are subtracted. The resulting error signal is fed to the ADC. Note, that the input stage operates over the entire voltage range of the analog interface ($\pm$\SI{10}{\volt}) which effectively extends the controller input voltage range without reducing the ADC resolution. Additional monitoring outputs for the amplified photodetector signal and the error signal are available. PyRPL is used to implement a PI driving a fast control output via an output stage. It consists of a buffer that can drive a \SI{50}{\ohm} load up to the increased maximal output voltage (\SI{2}{\volt}). An optional sample and hold feature is realized via PyRPL: One of the digital input/output pins (DIO) sets the PI gains to zero when pulled high, resulting in a constant control output after a delay of \SI{150(10)}{\nano\second}.

\begin{figure}[htb]
%	\centering
	\includegraphics[width = \columnwidth]{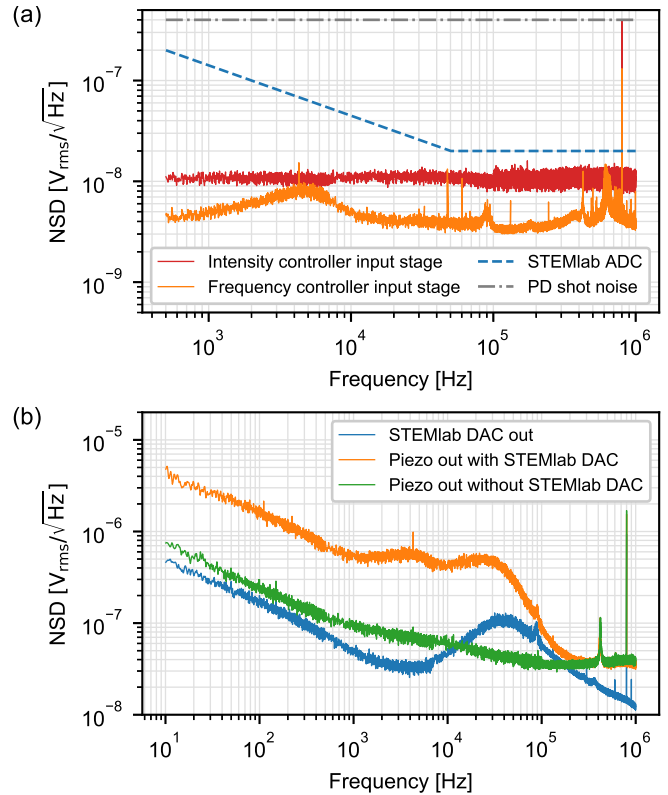}
	\caption{\label{fig:noise}Voltage noise spectral density (NSD) for the input stages (a) and output stages (b) of  Fig.~\ref{fig:block}. In (a), the STEMlab 125-14 ADC noise \cite{Olaya2017} and the shot noise level of a typical photodetector (PD) are plotted for comparison (see text). }
\end{figure}

\section{Noise and bandwidth performance}
For characterization of the performance of the controllers, a comprehensive analysis of the noise spectral density (NSD) of all relevant input and output stages is presented in Fig.~\ref{fig:noise}. All measurements have been performed with a low-noise pre-amplifier in combination with a high-resolution digital multimeter for frequencies below \SI{500}{\hertz}, a fast oscilloscope for frequencies between \SI{500}{\hertz} and \SI{100}{\kilo\hertz}, and a real-time spectrum analyser for frequencies above \SI{100}{\kilo\hertz}. Figure~\ref{fig:noise}(a) reveals a noise floor well below the ADC input noise\cite{Olaya2017} for the input stages of both controllers. The shot noise level of a typical photodetector operated with a photocurrent of \SI{200}{\micro\ampere} and a transimpedance gain of \SI{50}{\kilo\ohm} is plotted for comparison. This shows that the input NSD for the intensity controller is typically dominated by the shot noise of the detector. The above characterization has been carried out with the gains of the frequency error signal and the photodetector input set to unity. An upper bound of the NSD for larger gains can be obtained by multiplying the respective spectrum with the selected value of the gain.

Figure~\ref{fig:noise}(b) shows the NSD of all relevant output signals. The spectra can be used to determine the noise contribution of the controllers within the control loop.  The depicted NSD of the STEMlab DAC output (blue) also represents the fast output of the frequency controller (Fig.~\ref{fig:block}(a)) and the output of the intensity controller (Fig.~\ref{fig:block}(b)) since the contribution of the output buffer to the NSD is negligible. The broad noise peak around \SI{40}{\kilo\hertz} is an inherent feature of the STEMlab system and is not influenced by the analog interface hardware presented in this work. A  measurement of the NSD of the DAC output over its full bandwidth of \SI{50}{\mega\hertz} (not shown) yields an accumulated RMS noise of \SI{125}{\micro\volt}. The NSD of the STEMlab-controlled piezo output for frequency stabilization is represented by the orange spectrum. The accumulated RMS noise (\SI{10}{\hertz} to \SI{1}{\mega\hertz}) of this spectrum amounts to \SI{122}{\micro\volt}. For comparison, the NSD of the piezo stage disconnected from the STEMlab output is shown in green. This measurement shows that the NSD of the piezo output is dominated by the amplified NSD of the STEMlab system. Note, that the step-down regulator used for power supply produces a peak at its switching frequency (\SI{800}{\kilo\hertz}
\footnote{The switching frequency can be varied between \SI{0.2}{\mega\hertz} and \SI{2.2}{\mega\hertz}.}) clearly visible in all presented spectra. The RMS amplitude of this feature is below \SI{15}{\micro\volt} and thus not detectable by the ADCs that feature an effective resolution of 11.8 bits \cite{Olaya2017} corresponding to an RMS noise of \SI{162}{\micro\volt}. The switching peak's contribution to the STEMlab's output RMS noise is also negligible.

Both input stages, all monitoring outputs in Fig.~\ref{fig:block}, as well as the output buffer in Fig.~\ref{fig:block}(b) provide a \SI{-3}{\decibel}-bandwidth beyond \SI{5}{\mega\hertz} well above the maximally available control bandwidth of \SI{1.25}{\mega\hertz}. Thus, input as well as output characteristics of both types of controllers are limited solemnly by the performance of the STEMlab hardware concerning noise and bandwidth.

\begin{figure}[htb]
	\includegraphics[scale =1.1]{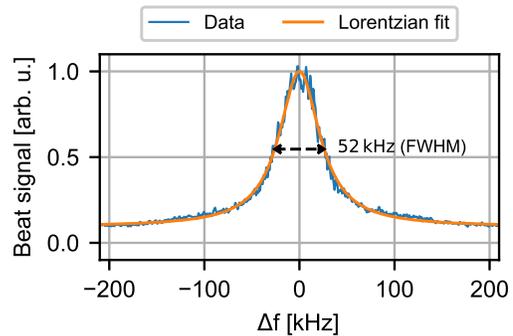}
	\caption{\label{fig:beat}Centred beat spectrum of two independently stabilized ECDL systems.}
\end{figure}

\section{Applications}
The frequency controller depicted in Fig.~\ref{fig:block}(a) is used to stabilize the frequency of the emitted light of an ECDL system to a spectroscopic feature of the D${}_2$ line of rubidium vapor at a wavelength of \SI{780}{\nano\meter}. The error signal is obtained by optimized  modulation transfer spectroscopy (MTS).\cite{Preuschoff2018} For this system, the control loop bandwidth is limited by the MTS error signal bandwidth of \SI{750}{\kilo\hertz}. In order to minimize the influence of DAC quantization noise, the control range of the fast actuator (i.e. laser current) is restricted to \SI{140}{\mega\hertz} by sufficient attenuation of the fast output. For the same purpose, an external low-pass filter (cut-off frequency \SI{200}{\hertz}) is applied to the laser piezo. Beating the outputs of the stabilized ECDL and an ECDL system stabilized independently to a high-finesse cavity yields an upper bound for the effective laser linewidth. Figure~\ref{fig:beat} shows the centred beat spectrum of the two laser systems averaged over approximately \SI{15}{\second}. Optimal performance is obtained for a P-gain of \SI{0.1}{} and a PI corner frequency of \SI{180}{\kilo\hertz} for the fast actuator. For these parameters, the unity gain frequency is given by the error signal bandwidth (\SI{750}{\kilo\hertz}). The line shape is predominantly Lorentzian with a full width at half maximum (FWHM) of \SI{52(1)}{\kilo\hertz}. A Lorentzian line shape indicates a frequency noise spectrum dominated by non-technical white noise\cite{Agrawal1988} with sufficient suppression of technical frequency noise by the active control loop. The digital controller presented in this work reduces the effective linewidth by a factor of three as compared to our previous work based on analog controllers.\cite{Preuschoff2018}

\begin{figure}[htb]
	\includegraphics[scale =1.1]{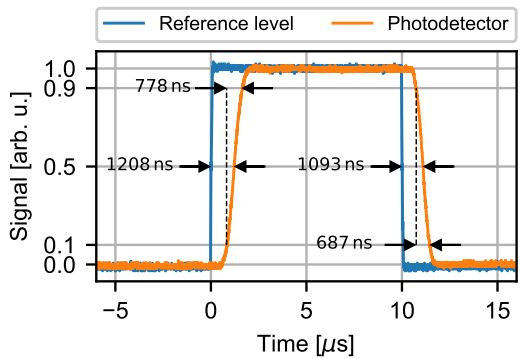}
	\caption{\label{fig:pulse}Light pulse generated by an AOM based intensity control loop (orange) and the applied control signal (blue).}
\end{figure}
An intensity control loop for laser light at \SI{780}{\nano\meter} wavelength is implemented using the intensity controller shown in Fig.~\ref{fig:block}(b). The controller regulates the deflection efficiency of an AOM by controlling the RF power driving the AOM. This is achieved by applying the control output to the intermediate port of a frequency mixer (Mini Circuits ZAD-3+). Note, that the control loop gain depends on the intensity setpoint in relation to the maximally deflected intensity due to the non-linearity of the frequency mixer. The laser intensity is measured by a fast photodetector connected to the controller input. The photodetector consists of a silicon photodiode (Osram SFH203 FA) and a \SI{50}{\kilo\ohm}-gain transimpedance amplifier with a \SI{-3}{\decibel}-bandwidth of \SI{1.8}{\mega\hertz}. The transit time of the acoustic wave in the AOM crystal induces a significant delay in the system limiting the available control bandwidth. The PI corner frequency is set to \SI{1.25}{\mega\hertz} and the P-gain is set to 0.02. For these parameters, we infer a unity gain frequency of \SI{170}{\kilo\hertz}. Figure~\ref{fig:pulse} illustrates the performance of the intensity control loop. A \SI{10}{\micro\second} light-pulse is generated by applying an appropriate voltage pulse to the controller's reference-level input (blue). The pulse height corresponds to \SI{90}{\percent} of the maximally deflected laser power. The resulting light pulse (orange) is recorded with an independent photodetector.  We retrieve a rise (fall) time of \SI{778(15)}{\nano\second} (\SI{687(11)}{\nano\second}). The delay of \SI{1208}{\nano\second} (\SI{1093}{\nano\second}) between the reference signal and the intensity pulse for the rising (falling) edge exhibits a low RMS jitter of \SI{7}{\nano\second} (\SI{9}{\nano\second}). Analysis of the out-off-loop photodetector signal for a constant reference-level input shows that the relative RMS intensity noise is suppressed to a level of $\Delta I_{rms} / I <$ \SI{1e-3}{} in the frequency range \SIrange{0.1}{1e5}{\hertz}. For frequencies below \SI{300}{\hertz} the relative NSD shows 1/$\sqrt{\mathrm{f}}$-behavior at a level below \SI{2e-5}{}.

\section{Conclusion}
In conclusion, the hardware design presented in this work comprises all input and output stages necessary for laser frequency and intensity stabilization in compact and cost efficient devices based on the STEMlab 125-14 board. For the resulting controllers, the noise performance and bandwidth are not influenced by the presented hardware making the full potential of the STEMlab platform available. The wide range and high reproducibility of the applicable digital PI control parameters allow for the application of the presented devices to various control tasks, beyond the two examples presented in this work, without any hardware modifications.  However, if modifications are necessary, the open-source availability of hardware and software offers additional flexibility.\\

\section{Acknowledgments}
This work has been supported by the Deutsche Forschungsgemeinschaft (DFG) [Grant BI 647/6-2, Priority Program SPP 1929 (GiRyd)] and the BMBF [Grant 05P19RDFAA]. We thank L. Neuhaus for support with the PyRPL software package and P. Baus for helpful discussions and advice.\\

The data that support the findings of this study are openly available in the GitHub repository TU-Darmstadt-APQ.\cite{APQGit_Lockbox,APQGit_IntStab}
\bibliography{RedPitaya_arXiv}% Produces the bibliography via BibTeX.
\end{document}